\documentclass[a4paper,12pt]{article}
\pdfoutput=1 
\usepackage{jheppub} 
\usepackage[utf8]{inputenc}
\usepackage{amsthm}
\usepackage{mathrsfs} 
\usepackage[all]{xy}
\graphicspath{{./images/}}
\theoremstyle{plain}

\theoremstyle{definition}

\numberwithin{thm}{section}

\newcommand{\bra}[1]{ \langle {#1} | }
\newcommand{\ket}[1]{ | {#1} \rangle }

\def\d{{\rm d}}
\def\i{{\mathsf i}}

\def\sign{\mathop{\rm sign}\nolimits}

\usepackage{slashed}

\def\cJ{{\cal J}}

\def\cL{{\cal L}}

\def\bR{{\mathbb R}}

\def\bZ{{\mathbb Z}}

\def\sA{{\mathsf A}}

\def\sC{{\mathsf C}}

\def\sF{{\mathsf F}}

\def\sJ{{\mathsf J}}

\def\sa{{\mathsf a}}

\def\sc{{\mathsf c}}

\def\U{\mathrm{U}}
\def\SU{\mathrm{SU}}

\def\beq#1\eeq{\begin{align}#1\end{align}}

\def\w{\widetilde}
\def\h{\widehat}
\def\sphi{\mathsf \Phi}

\title{Witten effect, anomaly inflow, and charge teleportation }

\preprint{TU-1107}

\author[1,2]{Hajime Fukuda}
\author[3]{and Kazuya Yonekura}
\affiliation[1]{
Theoretical Physics Group, Lawrence Berkeley National Laboratory, CA 94720, USA
}
\affiliation[2]{
Berkeley Center for Theoretical Physics, Department of Physics,\\
University of California, Berkeley, CA 94720, USA
}
\affiliation[3]{
Department of Physics, Tohoku University, Sendai 980-8578, Japan
}

\abstract{ 
We study a phenomenon
that electric charges are ``teleported'' between two spatially separated objects without exchanging charged particles at all.
For example, this phenomenon happens  between a magnetic monopole and an axion string in four dimensions, two vortices in three dimensions,
and two M5-branes in M-theory in which M2-charges are teleported. 
This is realized by anomaly inflow into these objects in the presence of cubic Chern-Simons terms.
In particular, the Witten effect on magnetic monopoles can be understood as a general consequence of anomaly inflow,
which implies that some anomalous quantum mechanics must live on them. 
Charge violation occurs in the anomalous theories living on these objects, 
but it happens in such a way that the total charge is conserved between the two spatially separated objects.
We derive a formula for the amount of the charge which is teleported between the two objects in terms of the linking number of their world volumes in spacetime. 
} 

\begin{document}

\maketitle

\section{Introduction} 

The Witten effect~\cite{Witten:1979ey} says that magnetic monopoles and dyons have electric charges of the form $\theta/2\pi + n$,
where $\theta$ is the $\theta$ angle of the $\U(1)$ gauge field and $n \in \bZ$ is an integer. 
It leads to a curious phenomenon when there exist axion strings.\footnote{
See also \cite{Hidaka:2020iaz,Hidaka:2020izy} for a recent work where this system has been studied from a different perspective. 
}

An axion field $\sphi$ can be coupled to the $\U(1)$ gauge field by the usual topological term which is proportional to $\sphi \sF \wedge \sF$,
where $\sF = \frac{1}{2} \sF_{\mu\nu} \d x^\mu \wedge \d x^\nu$ is the $\U(1)$ field strength. 
In a normalization of $\sphi$ such that the $\theta$ angle is given by $\theta = 2\pi \sphi$, monopoles and dyons have electric charges $\sphi + n$.
An axion string is a codimension 2 object which may be defined by the property that the vacuum expectation value of $\sphi$ changes as $\sphi \to \sphi +1$
when we go around it. 

\begin{figure}
\centering
\includegraphics[width=.5\textwidth]{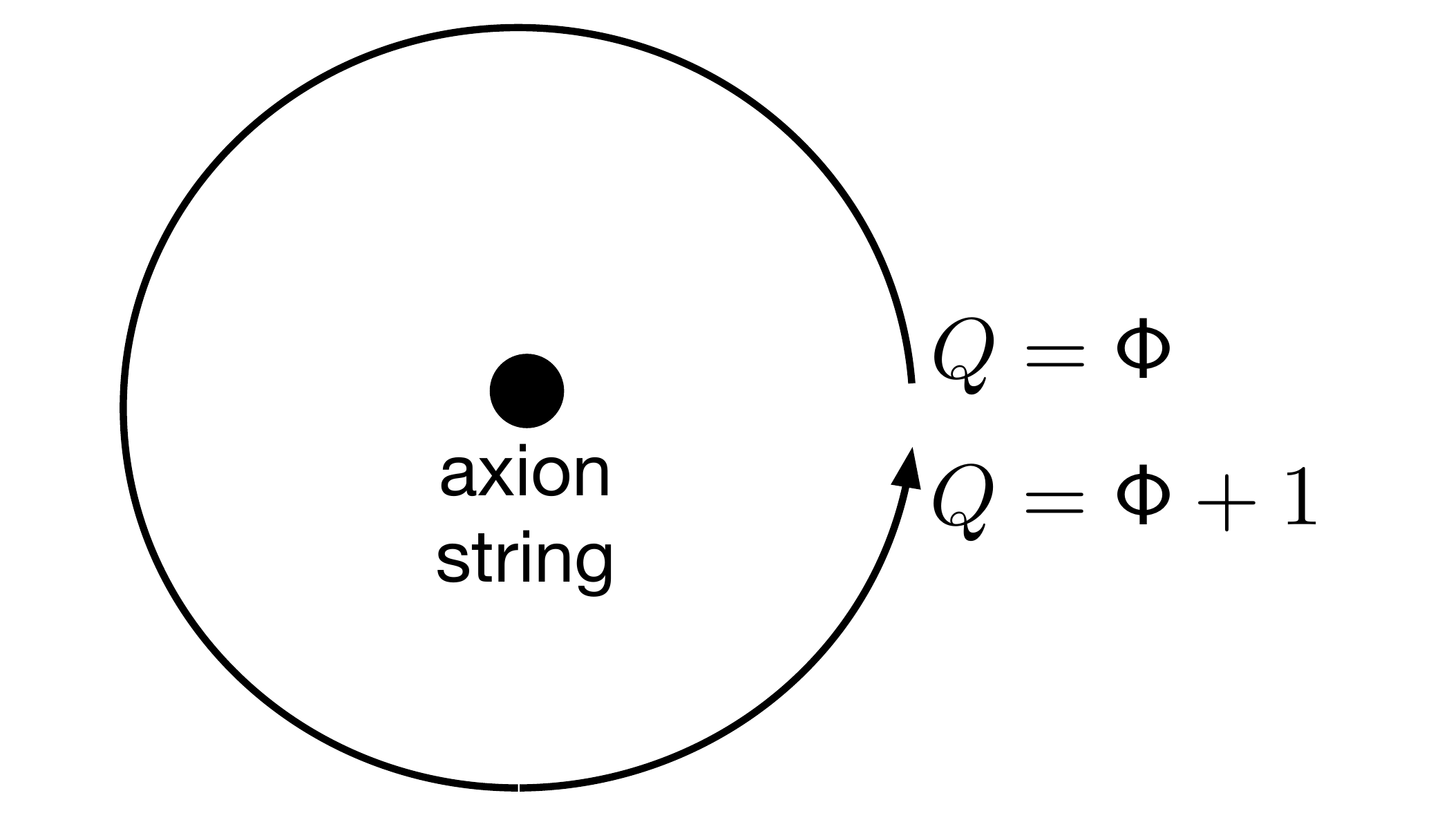}
\caption{ An axion string is characterized by the property that $\sphi$ has monodromy $\sphi \to \sphi +1$ around it. 
The electric charge of a magnetic monopole, which we denote as $Q$, changes after going around an axion string. \label{fig:1}}
\end{figure}

Now suppose that we have a magnetic monopole in our hand, and there is an axion string in the universe.
We travel by a spacecraft and go around the axion string, and then return to the earth. 
During the travel, the value of $\sphi$ which the monopole feels is gradually changed from some value $\sphi= \sphi_{\rm earth}$ to 
another value $\sphi = \sphi_{\rm earth}+1$. This means that the value of the electric charge of the monopole
is changed by a unit charge. See Fig.~\ref{fig:1}. If we repeat this process, we can produce electric charges without any limit.
We can assume that the spacecraft is shielded so that there is no exchange of charged particles between the regions inside and outside the spacecraft.
Then, naively, it seems that charge conservation is violated. However, charge conservation is
one of the most important requirements from the consistency of $\U(1)$ gauge theory. 

The resolution of the above puzzle is as follows. The magnetic monopole produces the usual magnetic field $\sF$,
and the axion string feels it. Due to anomaly inflow into the axion string~\cite{Callan:1984sa},
there must be some chiral fermions on the axion string which are anomalous under the $\U(1)$ gauge symmetry.
Then, the $\U(1)$ charge is not conserved also on the axion string due to the effect that the vacuum $\U(1)$ charge of an anomalous chiral fermion
in 2 dimensions depends on the background $\U(1)$ gauge field $\sA$ in a continuous way. See Appendix~\ref{app:chiral} for a review.
By the motion of the monopole, the gauge field $\sA$ on the axion string is varied and
$\U(1)$ charge is excited on the axion string in such a way that the total electric charge
on the monopole and the axion string is conserved.
The situation is as if the electric charge is ``teleported'' between the monopole and the axion string.
Thus we call this phenomenon as {\it charge teleportation}. Of course, there is no inconsistency with causality
since the information of the scalar expectation value $\sphi$ and the magnetic flux $\sF$ cannot travel faster than the speed of light.
However, there is no exchange of charged particles like electrons between them. 
We can understand this phenomenon as anomalous generation of electric charges on each of the magnetic monopole and the axion string
in a consistent way so that the total charge is conserved. 

It is not restricted to the system of a monopole and an axion string in 4 dimensions.
Another concrete example is a system of two M5-branes in M-theory.
We can move two M5-branes while keeping finite distance between them.
Then, depending on the topology of their motion, M2-brane charges are generated on each of the M5-branes.
Thus it is a teleportation of M2-charges between two M5-branes without actually exchanging any M2-brane.

In this paper, we study the above phenomenon in a systematic way
in any dimension and for any magnetically charged objects.
We identify that the origin of the phenomenon is due to cubic Chern-Simons interactions such as $\sphi \sF \wedge \sF$,
where $\sphi$ is regarded as a 0-form gauge field with $(-1)$-form gauge symmetry $\sphi \to \sphi+1$.
In section~\ref{sec:cubic}, we discuss anomaly inflow into magnetically charged objects in the presence of cubic Chern-Simons interactions.
In section~\ref{sec:charge}, we explain how anomalies lead to charge generation on magnetic objects.
In particular, we show that the Witten effect can be understood as a general consequence of anomaly inflow,
where the relevant anomaly is that of $(-1)$-form gauge symmetry studied in~\cite{Cordova:2019jnf}.
In section~\ref{sec:teleportation}, after discussing a simple example of charge teleportation between two vortices in three spacetime dimensions,
we derive a general charge teleportation formula in terms of the linking number between the world volumes of two magnetically charged objects.
The linking number has appeared in the context of Chern-Simons gauge theories~\cite{Polyakov:1988md,Witten:1988hf,Horowitz:1989km}
where they determine Aharonov-Bohm phases of anyons and their generalizations. 
In our case, the effect is more dramatic in the sense that not only the phase of state vectors of an object is changed, but also 
the charge is changed.

\section{Anomaly inflow with cubic Chern-Simons interaction } \label{sec:cubic}

\subsection{The general theory}
We consider the following theory in $d$~dimensions. Suppose that there are $p_i$-form gauge fields $ \sA_i$,
where $i $ is an index labeling these gauge fields. In this theory, we consider the action
\beq
S = - 2\pi \int \frac{ 1 }{2 } h_{ij} \d \sA^i \wedge \star \d \sA^j +  2\pi \int \frac{1}{3! } \kappa_{ijk} \sA^i \wedge \d \sA^j \wedge \d \sA^k. \label{eq:action}
\eeq
Here $\star$ is the Hodge star which is more explicitly given by 
\beq
(\star \omega)_{\nu_1 \cdots \nu_{d-q}} = \frac{1}{q!} \omega_{\mu_1 \cdots \mu_q} \epsilon^{\mu_1 \cdots \mu_q}_{\qquad ~  \nu_1 \cdots \nu_{d-q}}   ,\qquad (\omega \text{ : $q$-form}).
\eeq
$h_{ij}$ and $\kappa_{ijk}$ are coefficients with the symmetry properties $h_{ij} = h_{ji}$ and
\beq
\kappa_{ijk} = (-1)^{(p_i+1)(p_j+1)} \kappa_{jik} = (-1)^{(p_j+1)(p_k+1)} \kappa_{ikj}.
\eeq
The symmetry property of $\kappa_{ijk}$ is most easily seen by noticing that the Chern-Simons term can be defined by
the integral of $\frac{1}{3! } \kappa_{ijk} \sF^i \wedge \sF^j \wedge \sF^k$ (where $\sF^i = \d \sA^i)$ over a manifold with one higher dimension whose boundary is the spacetime.
The sum over repeated indices are implicit. 

Roughly, $h_{ij}$ are inverse squared of electric couplings, and $\kappa_{ijk}$ are Chern-Simons levels which need to be quantized to be integers.
The fact that $\kappa_{ijk}$ need to be integers will become apparent by later discussions. 
Even if $\kappa_{ijk}$ are integers, it is still a nontrivial question whether the Chern-Simons term is well-defined or not at the nonperturbative level.
In this paper, we simply assume that we consider $d$, $p_i$ and $\kappa_{ijk}$ for which it is well-defined possibly after some modification;
see \cite{Witten:1996md} for the case of the cubic Chern-Simons term in M-theory.

We normalize the gauge fields in the following way.
Suppose that there is an object (which we call an electric brane) which has the unit electric charge under $\sA^i$.
Such a brane has the world volume dimension $p_i$.
Then the coupling of the brane to $\sA^i$ is given by
\beq
S \supset 2\pi \int_{M} \sA^i.
\eeq
where the integral is over the world volume $M$.
This normalization of $\sA^i$ may be different from the more standard one by a factor of $2\pi$ (but it may be standard in string theory 
in the unit $2\pi \sqrt{\alpha'}=1$). 
If we take a local coordinate system $(y^0, \cdots, y^{d-1})$ in which the world volume $M$ is given by
$y^{p_i} = \cdots = y^{d-1} =0$, we can define the delta function localized on it,
\beq
\delta_M = \delta(y^{p_i} ) \d y^{p_i} \wedge \cdots \wedge \delta(y^{d-1}) \d y^{d-1}.
\eeq
Then we can rewrite $\int_{M} \sA^i = \int \sA^i \wedge \delta_M$.
The equation of motion in the presence of the brane is given by
\beq
(-1)^{p_i+1} \d (h_{ij} \star \sF^j ) - \frac{1}{2 } \kappa_{ijk} \sF^j \wedge  \sF^k = \delta_M . \label{eq:EOM}
\eeq

Suppose that there is an object which has a magnetic charge under $\sA^i$. We call such an object as a magnetic brane.
The world volume dimension is $d-p_i-2$.
We take a sphere $S^{p_i +1}$ which surrounds the magnetic brane. Then the standard Dirac quantization argument (which we will briefly review below)
implies that the integral of $\sF^i $ is
\beq
\int_{S^{p_i +1}} \sF^i  \in \bZ,
\eeq
where $\bZ$ is the set of integers.

On a magnetic brane, there is anomaly inflow~\cite{Callan:1984sa} due to the existence of the Chern-Simons term.
Here we explain it following the discussions in \cite{Witten:1999eg,Witten:2016cio,Tachikawa:2018njr,Hsieh:2020jpj}. 
The explanations there can be extended to nonperturbative (global) anomalies. 
(See \cite{Hsieh:2020jpj} for concrete examples of nonperturbative anomaly inflow with a cubic Chern-Simons term of the above type in M-theory.)
But for simplicity, we discuss anomaly inflow only at the perturbative level in this paper.
We remark that all fields in the following discussions are supposed to be on-shell (i.e. satisfy equations of motion). 

We define the magnetic dual field strength as 
\beq
\w \sF_i = (-1)^{p_i+1}h_{ij} \star \sF^j .
\eeq
The point of this definition is as follows. First, it is gauge invariant. Next,
if $\kappa_{ijk}=0$, its integral over a sphere $S^{d-p_i-1}$ surrounding an electric brane is 
\beq
\int_{S^{d-p_i -1}} \w \sF_i = 1
\eeq
where we have used \eqref{eq:EOM} with $\kappa_{ijk}=0$.
Therefore, an electric brane looks like a magnetic object from the point of view of $\w \sF_i$.

\begin{figure}
\centering
\includegraphics[width=.4\textwidth]{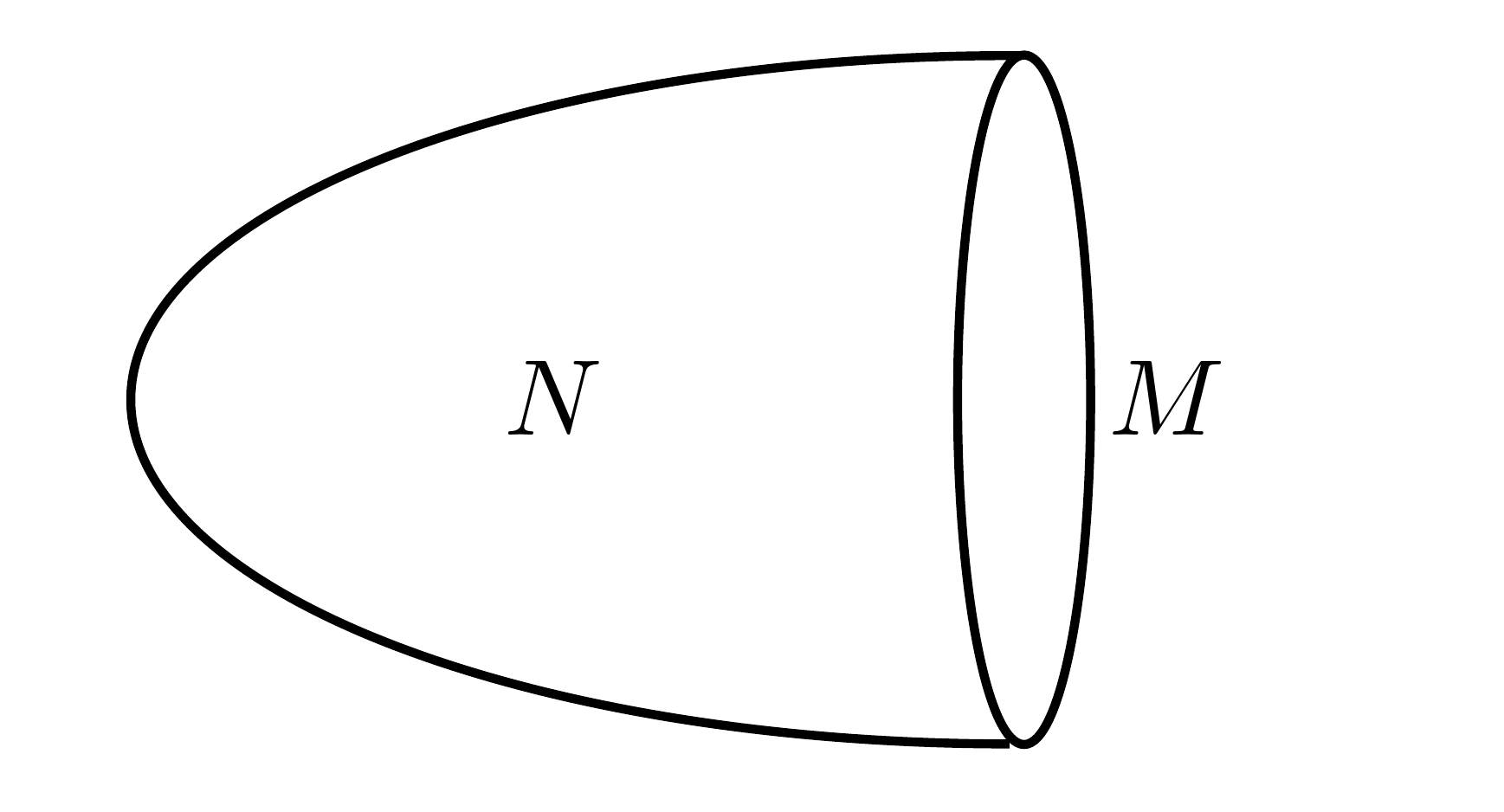}
\caption{ $M$ is the worldvolume of a magnetic brane, and we take a manifold $N$ whose boundary is $M$. \label{fig:2}}
\end{figure}

Schematically, we want to consider $\w \sF_i$ as $\w \sF_i \sim \d \w \sA_i$ for some dual gauge fields $\w \sA_i$.
The coupling of a magnetic brane with the world volume $M$ to $\w \sA_i$ is schematically given by
$2\pi \int_M \w \sA_i$. To define this coupling in terms of the original fields $\sA^i$ and $\sF^i$,
we introduce a submanifold $N$ such that its boundary is $M$, $\partial N = M$. See Fig.~\ref{fig:2}. Based on Stokes theorem,\footnote{
Given an orientation of $M$, the orientation of $N$ is defined by the condition that the Stokes theorem $\int_M \omega = \int_{N} \d \omega$ holds for differential forms $\omega$. 
} 
we expect that
\beq
2\pi \int_M \w \sA_i \to 2\pi \int_N \w \sF_i = 2\pi \int_N  (-1)^{p_i+1}h_{ij} \star \sF^j. \label{eq:coupling1}
\eeq
In this way, we can define the coupling of the magnetic brane to the gauge fields by using the original fields.
As we discuss below, this description is consistent as far as $\kappa_{ijk}=0$.
However, there will be a problem in the case of $\kappa_{ijk} \neq 0$.

Although the above description is gauge invariant, it may depend on a choice of $N$.
We want \eqref{eq:coupling1} to depend only on the world volume $M$ and not on a choice of $N$ such that $\partial N = M$.
To study its dependence on $N$, let us take another submanifold $N'$ such that $\partial N' = M$.
Then the difference of the coupling \eqref{eq:coupling1} between the cases of using $N$ and $N'$ is determined as follows.
$N$ and $N'$ have the common boundary $M$, and hence we can glue them along $M$ after flipping the orientation of $N'$.
Let $N_{\rm closed}$ be the manifold without boundary which is obtained by gluing $N$ and the orientation reversal of $N'$.
Then we get
\beq
 \int_N \w \sF_i  -  \int_{N'} \w  \sF_i  =  \int_{N_{\rm closed}} \w \sF_i .
\eeq
The action needs to be defined only modulo $2\pi$. If $\int_{N_{\rm closed}} \w \sF_i$ takes values in integers for any submanifold $N_{\rm closed}$ without boundary,
the coupling \eqref{eq:coupling1} is well-defined modulo $2\pi$. As long as $\kappa_{ijk}=0$, this requirement is consistent 
with the equation of motion \eqref{eq:EOM}. This is the standard argument for the Dirac quantization condition, but now with the roles of electric and magnetic 
fields exchanged. The quantization of $ \int_{N_{\rm closed}} \w \sF_i$ is also valid on manifolds with nontrivial topology.\footnote{
When $\kappa_{ijk}=0$, the relation between $\sF^i$ and $\w \sF_i$ are described roughly as follows.
We consider an action
\beq
S  = - 2\pi \int \frac{ 1 }{2 } h^{ij} \w \sF_i \wedge \star \w \sF_j - 2\pi (-1)^{(p_i+1)(d-p_i)} \int \w \sF_{i} \wedge \d \sA^{i},
\eeq
where $h^{ij}$ is the inverse matrix to $h_{ij}$. Here, we take the independent variables of the path integral to be $\w \sF_i$ and $\sA^i$.
If we integrate out $\w \sF_i$, one can check that we recover \eqref{eq:action} with $\kappa_{ijk}=0$. The on-shell value of $\w \sF_i$ is given
by $h^{ij}  \star \w \sF_j = - (-1)^{(p_i+1)(d-p_i)}  \d \sA^{i}$ which is equivalent to $\w \sF_i = (-1)^{p_i+1} h_{ij} \star \sF^j$ by using the property of Hodge star 
$\star^2 \w \sF_i = - (-1)^{(p_i+1)(d-p_i-1)} \w \sF_i$
in Minkowski signature metric. Alternatively, we can first integrate out $\sA^i$ which plays a role of Lagrange multiplier in the above action. 
First, by integrating over topologically trivial $\sA^i$, one gets the Bianchi identity $\d \w \sF_i=0$. 
We also need to sum over topologically nontrivial $\sA^i$ such that the integrals of $\d \sA^i$ over closed cycles are integers.
Then we find that the flux of $\w \sF_i$ is also quantized essentially due to the identity $\sum_{m \in \bZ } e^{2\pi \i m x} = \sum_{n \in \bZ} \delta( x -n)$ and Poincare duality.
See \cite{Witten:1995gf} for more detailed discussions.
}

But in the presence of nonzero $\kappa_{ijk}$, we have a modified Bianchi identity of $\w \sF_i$ given by
\beq
\d \w \sF_{i} = \frac{1}{2} \kappa_{ijk} \sF^j \wedge \sF^k,
\eeq
where $ \w \sF_i =   (-1)^{p_i+1}h_{ij} \star \sF^j$, and we have neglected the delta function source term in \eqref{eq:EOM} which causes no problem for the present purposes.
Therefore, $\w \sF_i$ is not closed, $\d \w \sF_i \neq 0$.
For non-closed $\w \sF_i$, the integral $\int_{N_{\rm closed}} \w \sF_i$ changes continuously as we deform $N_{\rm closed}$.
However, if $\int_{N_{\rm closed}} \w \sF_i$ takes values in integers, it cannot change continuously under deformation of $N_{\rm closed}$. 
Thus we conclude that $\int_{N_{\rm closed}} \w \sF_i$ is not integer for $\kappa_{ijk} \neq 0$ and \eqref{eq:coupling1} depends on a choice of the artificial data $N$.

One may try to modify \eqref{eq:coupling1} by including an additional term as 
\beq
 2\pi \int_N  \left( \w \sF_i -  \frac{1}{2} \kappa_{ijk} \sA^j \wedge \d \sA^k \right). \label{eq:coupling2}
\eeq
Then the integrand is closed, $\d ( \w \sF_i -  \frac{1}{2} \kappa_{ijk} \sA^j \wedge \d \sA^k ) =0$.
However, it is no longer gauge invariant. First of all, the fields $\sA^i$ may not  be 
defined globally as differential forms on the spacetime manifold. In general, we can define them as differential forms
only locally, and we need to glue them together among patches by nontrivial gauge transformations.
It is nontrivial to make sense of \eqref{eq:coupling2} in such a situation; see \cite{Hsieh:2020jpj} for general discussions.

Even if we restrict ourselves to the situations in which $\sA^i$ are globally well-defined as differential forms, there is still a problem which will lead to anomaly inflow.
We can perform gauge transformations
\beq
\sA^i \to \sA^i + \beta^i,
\eeq
where $\beta^i$ are closed differential forms $\d \beta^i=0$ such that their integrals over $p_i$-cycles are integers, 
\beq
\int_{p_i\text{-cycle}} \beta^i \in \bZ.
\eeq
These gauge transformations include small gauge transformations $\beta^i = \d \alpha^i$
for which $\int_{p_i\text{-cycle}} \beta^i =0$, and large gauge transformations for which $\int_{p_i\text{-cycle}} \beta^i \neq 0$.
Then \eqref{eq:coupling2} changes by
\beq
 2\pi \int_N  \left(  -  \frac{1}{2} \kappa_{ijk} \beta^j \wedge \d \sA^k \right) =  2\pi  (-1)^{p_j+1} \int_M  \left(  \frac{1}{2} \kappa_{ijk} \beta^j \wedge  \sA^k \right). \label{eq:inflow}
\eeq
This means that there is inflow of anomaly due to the existence of the Chern-Simons interaction.
The world volume theory on the magnetic object must contain anomalous degrees of freedom which precisely cancel
the contribution from anomaly inflow to make the total system consistent.

\subsection{Example: axion string and magnetic monopole} \label{sec:monopole}

Here we consider a concrete example in $d=4$ dimensions.
Suppose that there is a 1-form $\U(1)$ gauge field $A = A_\mu \d x^\mu$ and an axion field $\phi$ (0-form) with periodicity $\phi \sim \phi +2\pi$.
We consider the coupling $\frac{\kappa}{8\pi^2} \int \phi F \wedge F$, where $F = \d A$.
To make the normalization consistent with the previous subsection, we define $\sA = A/2\pi$ and $\sphi = \phi/2\pi$
and consider 
\beq
 2\pi \int \frac{1}{2} \kappa  \sphi \d \sA \wedge \d \sA.  
\eeq
This means that we take $\kappa_{ijk}$ of the previous section to be $\kappa_{\sA\sA\sphi} =\kappa_{\sA\sphi\sA}=\kappa_{\sphi\sA\sA} =\kappa$.

In this theory, there are two types of magnetic objects.
One is axion strings and the other is magnetic monopoles.

First let us consider an axion string. Its defining property is that
if we go around the axion string, the field $\sphi =\phi/2\pi$ changes as $\sphi \to \sphi +1$. This in particular means that $\sphi$ is not single valued.
We can rephrase this definition as follows. Consider the field strength $\d \sphi$ of $\sphi$.
If we integrate $\d \sphi$ over a circle $S^1$ which goes around the axion string, the integral is 
\beq
\int_{S^1} \d \sphi =1.
\eeq
This means that the axion string has magnetic charge $1$ under the 0-form gauge field $\sphi$.
As we showed in the previous subsection, there is anomaly inflow from Chern-Simons of the form.
\beq
-2\pi \int_N \frac{1}{2} \kappa \sA \wedge \d \sA,
\eeq
where the integral is over a 3-dimensional manifold $N$ such that its boundary $\partial N$ is the world sheet of the axion string.
As is well known, it can be cancelled by the anomaly of chiral fermions (or chiral bosons) living on the string world sheet.

Next let us consider a magnetic monopole. Its defining property is that the integral of the field strength $ \d \sA$ over a sphere $S^2$
surrounding the monopole is given by
\beq
\int_{S^2} \d \sA =-1,
\eeq
where the sign is chosen to be consistent with the convention used in this paper 
and will be discussed later (see \eqref{eq:magneticsign}).
By the general argument of the previous subsection, we know that there is anomaly inflow into the monopole,
represented by
\beq
-2\pi  \int_{N} \kappa \sphi \d \sA, \label{eq:Manomaly}
\eeq
where the integral is over a 2-dimensional manifold $N$ such that its boundary $\partial N$ is the world line of the monopole.
Compared to the previous subsection, we have added a total derivative term $-\pi \kappa \d( \sphi \sA)$ which does not affect the argument of the previous subsection.
It can be written as a local action on the boundary $\partial N$ by using Stokes theorem, 
and it represents the well-known ambiguity of local counterterms in the presence of anomalies.

Let us consider the anomaly inflow into the monopole in a little more detail.
The field $\sphi$ is a zero form field and hence its ``gauge transformation''
might be slightly unfamiliar. There is no small gauge transformation. However,
there is large gauge transformation $\sphi \to \sphi + \beta$ for integers $\beta \in \bZ$.
Under this gauge transformation, \eqref{eq:Manomaly} changes by
\beq
-2\pi \kappa \int_{N} \beta \d \sA = -2\pi  \kappa \beta \int_{M} \sA. \label{eq:Manomaly2}
\eeq
where $M = \partial N$ is the world line of the monopole.
Therefore, there must be quantum mechanical degrees of freedom whose anomaly cancels this contribution.

One example of such a quantum mechanical model is as follows~\cite{Cordova:2019jnf}. (See also \cite{Gaiotto:2017yup,Kikuchi:2017pcp} 
for closely related discussions.)
We consider a quantum mechanical model described by $q(t)$ with the target space $S^1$, that is, $q \in S^1$.
The fact that it lives on $S^1$ can be realized by a periodicity $q \sim q+1$.
Moreover, we assume that it has a $\U(1)$ symmetry $q \to q + \kappa' \alpha$,
where $\alpha$ is a $\U(1)$ transformation parameter and $\kappa'$ is the $\U(1)$ charge of $e^{2\pi \i q}$.
Because of this symmetry,
we can couple this model to the $\U(1)$ gauge field $\sA$ by introducing the covariant derivative 
$  \dot{q } -  \kappa'  \sA_t$,
where $\dot{q}$ is the derivative of $q$ by the time coordinate $t$. We consider the action of the form
\beq
S = \int \d t \frac{m}{2}(\dot{q } -  \kappa' \sA_t)^2 - 2\pi \kappa'' \int \d t \sphi (\dot{q } -  \kappa' \sA_t) \label{eq:QMmodel}
\eeq
where $m$ and $\kappa''$ are parameters. 
We take $\kappa''$ to be an integer by the reason which will become clear below.
However, the detail of the kinetic term is not important at all and we can replace it by any function of $(\dot{q } -  \kappa' \sA_t)$
as $\frac{m}{2}(\dot{q } -  \kappa' \sA_t)^2 \to F(\dot{q } -  \kappa' \sA_t)$.

The important term for our purposes is the one proportional to $\sphi$. This term represents a coupling of the model to $\sphi$, and plays 
a similar role as the $\theta$ angle in four dimensional gauge theories.
Let us perform the gauge transformation $\sphi \to \sphi + \beta$ for $\beta \in \bZ$.
Then the change of this term is
\beq
-2\pi \beta \kappa''  \int \d t (\dot{q} - \kappa' \sA_t). \label{eq:Qchange}
\eeq
The first term which is proportional to $\int \d t\, \dot{q}$ has no effect by the following reason. Suppose that we are computing a transition amplitude
$\bra{q_f} e^{ - \i t H} \ket{q_i}$. This means that the initial condition is $q(t_i) = q_i$ and the final condition is $q(t_f) = q_f$.
Then the effect of the first term of \eqref{eq:Qchange} is
\beq
\exp \left( - 2\pi \i \kappa'' \beta \int^{t_f}_{t_i} \d t\, \dot{q} \right) = \exp({- 2\pi \i \kappa'' \beta q_f}) \exp({ 2\pi \i \kappa'' \beta q_i})
\eeq
This contribution can be eliminated by a simple redefinition of state vectors $\ket{q}$ by a phase factor as
\beq
\ket{q} \to \ket{q}' :=\exp(-{ 2\pi \i \kappa'' \beta q}) \ket{q}. \label{eq:basechange}
\eeq
Notice that $\exp(-{ 2\pi \i \kappa'' \beta q})$ is single valued on $S^1$ (i.e. invariant under $q \to q+1$) as far as $\kappa''$ is an integer.
This is the reason that we take $\kappa''$ to be an integer. 
By this redefinition, the first term of \eqref{eq:Qchange} is completely eliminated. 

However, \eqref{eq:Qchange} contains the term
\beq
2\pi \beta \kappa' \kappa''  \int \d t  \sA_t 
\eeq
which cannot be eliminated by a redefinition of $\ket{q}$.\footnote{
By $\U(1)$ gauge transformation $\ket{q} \to e^{2\pi \i \alpha(q) }\ket{q}$,
we can change $\sA_t$ as $\sA_t \to \sA_t + \partial_t \alpha$.
}
This is the anomaly of this quantum mechanical model. 
See \cite{Cordova:2019jnf} for more detailed discussions. 

Notice that the above anomaly is precisely the opposite to the inflow contribution \eqref{eq:Manomaly2}
if we take the parameters $\kappa'$ and $\kappa''$ to be such that $\kappa' \kappa'' = \kappa$.
Therefore, if this quantum mechanics is living on the world line of the monopole, the 
anomalies from the inflow and the quantum mechanical model cancel each other and the total system is consistent.

The world line theory need not be precisely the model \eqref{eq:QMmodel}. From the point of view of anomalies,
any model is allowed as far as it produces the same anomaly as above. However, the model based on the field $q(t)$ with target space $S^1$
seems to be quite good in known examples of monopoles. For example, in a 't~Hooft Polyakov monopole constructed from a configuration of $\SU(2)$
gauge theory coupled to a scalar in the adjoint representation, there are moduli of monopole solutions which are parametrized by 
$\bR^3 \times S^1$. The first factor $\bR^3$ is associated to the spontaneous breaking of translation symmetry by a monopole solution, or in other worlds the position
of the monopole. The factor $S^1$ is associated to the spontaneous breaking of the $\U(1)$ symmetry by a monopole solution. (See e.g. \cite{Harvey:1996ur}
for a review.)
We need to quantize these degrees of freedom associated to moduli, 
and the quantization of the modulus parametrizing $S^1$ leads to a model analogous to the above quantum mechanical model.
Another example of a monopole is obtained by compactification of some branes in string theory.
For example, we can consider M5-branes in M-theory.
The 11-dimensional M-theory can be compactified to 4 dimensions, and an M5-brane can be 
compactified to a particle in 4 dimensions which behaves as a monopole.
In that case, a brane contains a $p$-form gauge field (with $p=2$ in the case of an M5-brane) on the world volume and its compactification gives a 0-form gauge field on the monopole. This zero form gauge field is described by $q(t)$ with target space $S^1$.

In this subsection, we have focused on axion strings and monopoles.
It is possible to construct examples of anomalous theories for more general spacetime dimensions and world volume dimensions.
See \cite{Hsieh:2020jpj} for the construction of these anomalous theories.

\section{Charge generation by anomaly}\label{sec:charge}

\subsection{The general case}\label{sec:QonBrane}

Anomalies lead to violation of charge conservation. Let us discuss it in our context. 
In this subsection, we forget the $d$-dimensional spacetime, and instead consider
the following situation. All the gauge fields $\sA^j$ are treated as background (external) fields.
We consider a $(d-p_I-1)$-dimensional ``topological phase'' $N$
with boundary $M = \partial N$ as in Fig.~\ref{fig:2}. Here, a ``topological phase'' means a theory with no degrees of freedom,
and contains only the action
\beq
2\pi \int_N  \left( -  \frac{1}{2} \kappa_{Ijk} \sA^j \wedge \d \sA^k \right). \label{eq:topoaction}
\eeq
This is the same as \eqref{eq:coupling2} but without the term $\tilde \sF_i$ which plays no role for the consideration of anomalies. 
As mentioned above, we treat $\sA^j$ as background fields and hence this action is constructed from purely background fields.
We have in mind the situation that $M$ is a world volume of some magnetic brane,
and $\sA^j$ are background fields created by another magnetic brane.

The $(d-p_I -2)$-dimensional boundary $M$ has an anomalous theory so that the anomaly inflow from
the topological phase $N$ and the anomaly of the boundary theory cancel with each other.
Therefore, the total system is anomaly free. This fact is important in the following discussion.

We define the current $\cJ_{I, \, j}$ associated to the gauge field $\sA^j$ by the variation of the action with respect to $\sA^j$.
(The index $I$ in the subscript of $\cJ_{I, \, j}$ is irrelevant in the discussion of this section.)
We consider an infinitesimal change $\sA^j \to \sA^j + \delta \sA^j$ and the current is defined as a differential form which appears as
\beq
\delta S = 2\pi  \int_N \delta \sA^j \wedge \cJ_{I, \, j}.
\eeq
We emphasize that the current $\cJ_{I, \, j}$ of the total system including both the bulk topological phase and the boundary anomalous theory is gauge invariant
since the total system is gauge invariant. The gauge invariance also implies the current conservation
\beq
\d \cJ_{I, \, j} =0.
\eeq

First, let us consider the case that the boundary is empty, $\partial N = \varnothing$.
The current is easily obtained from \eqref{eq:topoaction} as 
\beq
\cJ_{I, \, j} =  - \kappa_{Ijk}   \sF^k.
\eeq
It satisfies $\d \cJ_{I, \, j} =0$. 

Next we include the effect of the boundary $M = \partial N \neq \varnothing$.
Near the boundary, $N$ is given by $(-\epsilon, 0] \times M$,
where $\epsilon$ is an arbitrary small positive number which specifies a neighborhood of the boundary. 
We take the coordinate of $(-\epsilon, 0]$ as $\tau$.
Then the current $\cJ_{I, \, j}$ in this neighborhood is of the form
\beq
\cJ_{I, \, j} = - \kappa_{Ijk}  \theta( -\tau)  \sF^k + (-1)^{p_j}  \delta(\tau)\d \tau  \wedge  \sJ_{I,\, j}  
\eeq
where $\theta(s)$ is the step function with $\theta(s) =1$ for $s>0$ and $\theta(s)=0$ for $s<0$,
and $\delta(\tau) \d \tau$ is the delta-function 1-form localized at the boundary $\tau =0$.
The factor $\sJ_{I,\, j}$ is the current of the boundary theory, since this expression of $\cJ_{I, \, j}$ gives
\beq
\int \delta \sA^j \wedge \cJ_{I, \, j} = \int_N \delta \sA^j \wedge ( - \kappa_{Ijk}   \sF^k) + \int_M \delta \sA^j \wedge \sJ_{I,\, j}, \label{eq:var}
\eeq
where the first term is interpreted as the contribution from the bulk topological phase on $N$ and the second term is the contribution from the boundary anomalous theory on $M$.
(We will make remarks on this point in a moment.)
As emphasized above, the total system is anomaly free and hence $\cJ_{I, \, j}$ must satisfy the conservation equation
$\d \cJ_{I, \, j} =0$. By noticing that $\d [ \theta( -\tau) ]= - \delta(\tau) \d \tau$, we see that $\d \cJ_{I, \, j} =0$ gives
\beq
\d  \sJ_{I,\, j}   =  (-1)^{p_j} \kappa_{Ijk}   \sF^k. \label{eq:nonC}
\eeq
This equation means that the charge conservation is violated in the boundary anomalous theory.
This is one of the consequences of the anomaly. 

Let us make some remarks about the current $\sJ_{I,\, j}$. 
First, in the above discussion, we have formulated the currents in a gauge invariant way. This have resulted in
the ``covariant anomaly''\,\cite{Bardeen:1984pm,Naculich:1987ci} for the current of the higher form gauge symmetry. 
When one tries to find the current $\sJ_{I,\, j}$ from variation of the anomalous theory, one needs to
be careful about the following point.
If we use a certain renormalization (or in other words a certain choice of counterterms), the anomaly of the boundary theory is the opposite to the contribution from the inflow \eqref{eq:inflow}.
For a gauge transformation $\delta \sA^j= \d \alpha^j$, this anomaly of the boundary theory is
\beq
2\pi   \int_M  \left(  \frac{1}{2} \kappa_{Ijk} \alpha^j \wedge  \sF^k \right).
\eeq
where we have taken $i \to I$.
In the same renormalization scheme, let us define a current $\sJ'_{I,\, j}$ as
\beq
\delta S = 2\pi \int_M \delta \sA^j \wedge \sJ'_{I,\, j}.
\eeq
By taking $\delta \sA^j = \d \alpha^j$ and comparing with the above anomaly, we obtain
\beq
\d \sJ'_{I,\, j} = \frac{1}{2}  (-1)^{p_j}  \kappa_{Ijk}   \sF^k.
\eeq
This is the ``consistent anomaly'' for the current of the $p_j$-form gauge field.
Naively it seems that we get a different amount of anomaly than the one in \eqref{eq:nonC}.
However, the problem is that in the renormalization scheme defined above, the current $\sJ'_{I,\, j}$ is not gauge invariant.
This statement may be directly checked by considering a one point function $\langle \sJ'_{I,\, j} \rangle $ in the presence of the background fields $\sA^k$
and perform its gauge transformation.
Instead, we give the following argument. Let us combine the boundary theory with the bulk \eqref{eq:topoaction}.
The total system is anomaly free. The variation $ \delta \sA^j$ gives
\beq
\frac{\delta S}{ 2\pi}   &=    \int_N  \delta\left( -  \frac{1}{2} \kappa_{Ijk} \sA^j \wedge \d \sA^k \right) + \int_M \delta \sA^j \wedge \sJ'_{I,\, j} \nonumber \\
&=  \int_N \delta \sA^j \wedge ( - \kappa_{Ijk}   \sF^i) + \int_M \delta \sA^j \wedge \left(  \sJ'_{I,\, j} + \frac{1}{2} (-1)^{p_j} \kappa_{Ijk} \sA^k \right)
\eeq
where the last term of the second line appears due to integration by parts in the integral 
$ \int_N  \left( -  \frac{1}{2} \kappa_{Ijk} \sA^j \wedge \d \delta \sA^k \right)$,
and we have used $\kappa_{Ijk} = (-1)^{(p_j+1)(p_k+1)}\kappa_{Ikj}$ and a change of dummy indices $ j \leftrightarrow k$. By comparing with \eqref{eq:var}, we identify
\beq
\sJ_{I,\, j} = \sJ'_{I,\, j} + \frac{1}{2} (-1)^{p_j} \kappa_{Ijk} \sA^k .
\eeq
This $\sJ_{I,\, j}$ indeed satisfies the equation \eqref{eq:nonC}.
The gauge invariance of the total system implies that the current $\sJ_{I,\, j}$ is gauge invariant. 
Then the current $\sJ'_{I,\, j}$ is not gauge invariant. $\sJ'_{I, j}$ and $\sJ_{I, j}$ is the consistent and covariant current for the $p_j$-form gauge symmetry, respectively.
We stress that the total bulk-boundary system is gauge invariant and hence it is better to consider only gauge invariant quantities. 
The current $\sJ_{I, j}$ is such a gauge invariant quantity. 

Second, although we have derived the anomaly by focusing solely on the anomaly structure of the worldvolume $M$ and its extension $N$, 
we could obtain the same formula from the original action with the Chern-Simons term, \eqref{eq:action}, as is originally done in \cite{Callan:1984sa}.
For computations, we first need to determine the sign of the magnetic field created by a magnetic brane.
Suppose we have a magnetic brane associated to $\sA_i$. If we have the coupling \eqref{eq:coupling1},
the equation of motion for $\w \sA_i$ (by regarding it as a fundamental variable rather than $\sA^i$) is given by
\beq
(-1)^{d - p_i -1} \d ( h^{ij}  \star \w \sF_j) = \delta_{M}
\eeq
where $h^{ij}$ is the inverse matrix of $h_{ij}$. This equation is the electromagnetic dual of \eqref{eq:EOM}.
By using $\w \sF_i = (-1)^{p_i+1} h_{ij} \star \sF^j$ and $\star^2  \sF^i = - (-1)^{(p_i+1)(d-p_i-1)}  \sF^i$ in Lorentz signature metric, we get
\beq
\d \sF^i = (-1)^{p_i(d-p_i)} \delta_{M} . \label{eq:magneticsign}
\eeq
This implies that the integral of $\sF^i$ over a sphere $S^{p_i+1}$ surrounding the magnetic brane is $ (-1)^{p_i(d-p_i)}$.
Given this preparation, we first write the variation of the action in terms of current operators under the infinitesimal change of the gauge field $\sA^j$, $\sA^j \to \sA^j + \delta \sA^j$ as
\begin{eqnarray}
\frac{\delta S}{2\pi} = \int \delta \sA^j\wedge J_j + \int_{M_I} \delta \sA^j \wedge \sJ_{I, j} 
\end{eqnarray}
where we have assumed that the magnetic brane $M_I$ supports the non-trivial current $\sJ_{I, j}$ and explicitly write the variation for $M_I$. Here, the bulk current $J_j$ is given as
\begin{eqnarray}
\label{eq:CSCurrent}
J_j = \frac{1}{2}\kappa_{jki} \sF^k\wedge \sF^i,
\end{eqnarray}
where we have ignored the kinetic term.
The total current $\cJ_j$ is given by\footnote{
We require that the orientation of $M_I$ and the sign factor of $\delta_{M_I}$ are correlated as
$\int_{M_I} \omega = \int \omega \wedge \delta_{M_I}$ for arbitrary $\omega$.
}
\begin{eqnarray}
\cJ_j \equiv \frac{1}{2\pi}\frac{\delta S}{\delta \sA^j} = J_j + \sJ_{I, j} \wedge \delta_{M_I}.
\end{eqnarray}
It is gauge invariant and conserved, so that
\begin{eqnarray}
\d\cJ_j = 0 \iff \d\sJ_{I, j} \wedge \delta_{M_I} = - \d J_j.
\end{eqnarray}
Using \eqref{eq:magneticsign}, 
\begin{eqnarray}
-\d J_j = -(-1)^{p_k + 1} \kappa_{jkI} \sF^k\wedge (-1)^{p_I(d - p_I)}\delta_{M_I},
\end{eqnarray}
which results in $\d \sJ_{I, j} = (-1)^{p_j} \kappa_{Ijk} \sF^k$, the same equation as \eqref{eq:nonC}.
In this formalism, the gauge fields are dynamical. Thus, if $\kappa_{IIj} \ne 0$, for example, we need to be careful with the singularity of $\sJ_{I, j}$ due to $\sF^I$.
This singularity can give another anomaly~\cite{Freed:1998tg}, but for our purposes it is not necessary. 

Now let us consider a consequence of \eqref{eq:nonC}. The current $\sJ_{I,\, j}$ is a $p_k$-form.
To consider charges associated to $\sJ_{I,\, j}$ we take a $p_k$-dimensional cycle on $M$.
We assume that $M$ is of the form $\bR \times \h{M}$, where $\bR$ is a time direction parametrized by a time coordinate $t$,
and $\h{M}$ is space. Let us take a $p_k$-dimensional submanifold without boundary, $\Sigma \subset \h{M}$, and set $\Sigma_t = \{t\} \times \Sigma$.
We define charge $Q_{I,\, j}(\Sigma_t)$ as
\beq
Q_{I,\, j}(\Sigma_t) = \int_{\Sigma_t} \sJ_{I,\, j}.
\eeq
Comparing charges at $t = a$ and $t=b$, we get by Stokes theorem that
\beq
Q_{I,\, j}(\Sigma_b) - Q_{I,\, j}(\Sigma_a) &=  \int_{[a,b] \times \Sigma} \d \sJ_{I,\, j} \nonumber \\
&= (-1)^{p_j} \kappa_{Ijk}  \int_{[a,b] \times \Sigma}    \sF^k. \label{eq:QC1}
\eeq
This gives the charge non-conservation of the anomalous theory.
The right hand side can be rewritten again by Stokes theorem as
\beq
 \int_{[a,b] \times \Sigma}    \sF^k = \int_{\Sigma_b} \sA^k -  \int_{\Sigma_a} \sA^k
\eeq
Therefore the charge depends on $\sA^k$ as 
\beq
Q_{I,\, j}(\Sigma_t)= (-1)^{p_j} \kappa_{Ijk}  \int_{ \Sigma_t}    \sA^k  + \text{[state-dependent constant]},  \label{eq:QC3}
\eeq
where $\text{[state-dependent constant]}$ is a constant that depends on physical states but is independent of the time $t$.
See Appendix~\ref{app:chiral} for how this result is obtained in a direct computation in the case of chiral fermions in 2 dimensions.

If we include the bulk topological phase, the charge of the total system is conserved by gauge invariance.
The charge of the topological phase is ``vacuum charge'', since the topological phase does not have any degrees of freedom.

\subsection{Witten effect as a consequence of anomaly inflow}\label{sec:WQM}

As an example of the above phenomenon other than chiral fermions discussed in Appendix~\ref{app:chiral}, 
let us consider the magnetic monopole discussed in Sec.~\ref{sec:monopole}.
We want to consider the electric charge associated to the 1-form gauge field $\sA$.
In that case, we set $\kappa_{Ijk} \to \kappa,~p_j \to 1,~ \Sigma \to \,$point and $ \sA^k \to \sphi $.
By the above general discussions, the electric charge of the monopole is 
\beq
Q = - \kappa   \sphi  + \text{[const.]}, \label{eq:MEcharge}
\eeq
where $\text{[const.]}$ is independent of $\sphi$ (but depends on physical states).
This is a model-independent consequence of \eqref{eq:QC3} which comes from anomaly inflow.
The $\sphi$ is what is usually written in terms of the $\theta$ angle as $\sphi = \theta/2\pi$.
Therefore, we conclude that the Witten effect is a general consequence of anomaly inflow.

It is instructive to see how the charge non-conservation actually works in a concrete example.
Let us see it in the quantum mechanical model \eqref{eq:QMmodel}.

For simplicity we set $\sA=0$. But it is important to consider general nonzero values for $\sphi$.
Canonical quantization gives the canonical momentum $p$ and the Hamiltonian $H$ as
\beq
p &= m\dot{q} -2\pi \kappa'' \sphi, \\
H & = \frac{1}{2m} (p + 2\pi \kappa'' \sphi)^2.
\eeq
A naive charge associated to $\sA_t$ is given by $(\delta S/ \delta \sA_t)/2\pi = - \kappa' p/2\pi $.
However, $p$ is not gauge invariant under gauge transformations $\sphi \to \sphi + \beta$ ($\beta \in \bZ$) as the above expression of $p$ shows.
Thus we modify the current to get a gauge invariant expression
\beq
Q = - \frac{\kappa' m\dot{q} }{2\pi}= - \frac{\kappa' (p +2\pi \kappa'' \sphi)}{2\pi}.
\eeq

The physical states are obtained by noticing that the coordinate $q$ is periodic $q \sim q+1$ and hence wave functions are
spanned by $\psi_n(q) = \exp( 2\pi \i n q)$ for $ n \in \bZ$. The energy and charge of the state $\psi_n$ is 
\beq
E_n(\sphi) = \frac{4\pi^2}{2m} (n +  \kappa'' \sphi)^2, \qquad Q_n(\sphi) = - \kappa' n - \kappa \sphi,
\eeq
where $\kappa = \kappa' \kappa''$. The charge $Q_n(\sphi)$ is exactly as predicted by \eqref{eq:MEcharge},
and the state-dependent constant is explicitly given by $ - \kappa' n $.
The integer $n$ distinguishes various dyons. 

Now let us consider the situation where $\sphi$ is adiabatically changed from $\sphi $ to $\sphi +1$.
Notice that the spectrum of physical states is completely the same at $\sphi$ and $\sphi+1$, as expected from the gauge symmetry $\sphi \sim \sphi +1$.
However, by changing $\sphi \to \sphi+1$, the physical states $\psi_n$ are adiabatically shifted as $\psi_n \to \psi_{n+\kappa''}$, because
\beq
E_n(\sphi+1) = E_{n+\kappa''}(\sphi), \qquad Q_n(\sphi+1) = Q_{n+\kappa''}(\sphi).
\eeq
A reason behind it is as follows. The value $\sphi+1$ is equivalent to $\sphi$
if we perform gauge transformation $\sphi \to \sphi + 1$.
As discussed in Sec.~\ref{sec:monopole}, this gauge transformation must be accompanied by the change of basis vectors $\ket{q}$ as \eqref{eq:basechange},
\beq
\ket{q}' = \exp (- 2\pi \i \kappa'' q) \ket{q}.
\eeq
Therefore, the wave function becomes
\beq
\psi'_n(q) = \langle q |'  \psi_n \rangle = \exp (2\pi \i \kappa'' q) \langle q | \psi_n \rangle = \exp(2\pi \i \kappa'' q) \psi_n(q) = \psi_{n+\kappa''}(q).
\eeq
This explains the nontrivial monodromy $\psi_n \to \psi_{n+\kappa''}$ under the change of $\sphi$.
By this monodromy, the charge $Q$ is changed when $\sphi$ is adiabatically changed to $\sphi+1$.
One can check that essentially the same kind of monodromy happens in chiral fermions discussed in Appendix~\ref{app:chiral},

\section{Charge teleportation} \label{sec:teleportation}

\subsection{Example: two vortices in 3 dimensions}
Now we would like to discuss the phenomenon of charge teleportation mentioned in the Introduction.
A general discussion presented in the next subsection is technically a bit complicated,
so let us first see the phenomenon in the topologically simplest dimension, $d=3$.
We consider charge teleportation between two vortices.

This situation can be concretely realized by compactification of higher dimensional systems, such as $S^1$ compactification
of the system of a monopole and an axion string.
After the compactification, a monopole becomes a vortex of one kind, and an axion string wrapping $S^1$ becomes a vortex of another kind.
It would be very interesting to find a more realistic UV model or even actual material in condensed matter physics which realizes the IR effective theory of this subsection.
In this paper we just study effective theories with cubic Chern-Simons terms.

We consider a theory which contains two 0-form fields $\sphi_1$ and $\sphi_2$, and a 1-form field $\sA$.
The cubic Chern-Simons term is assumed to be given by
\beq
2\pi \int \kappa\sphi_1  \d \sphi_2 \wedge \d \sA.
\eeq
Magnetic branes associated to $\sphi_i~(i=1,2)$ are vortices around which $\sphi_i$ has a nontrivial monodromy $\sphi_i \to \sphi_i + 1$.

\begin{figure}
\centering
\includegraphics[width=.35\textwidth]{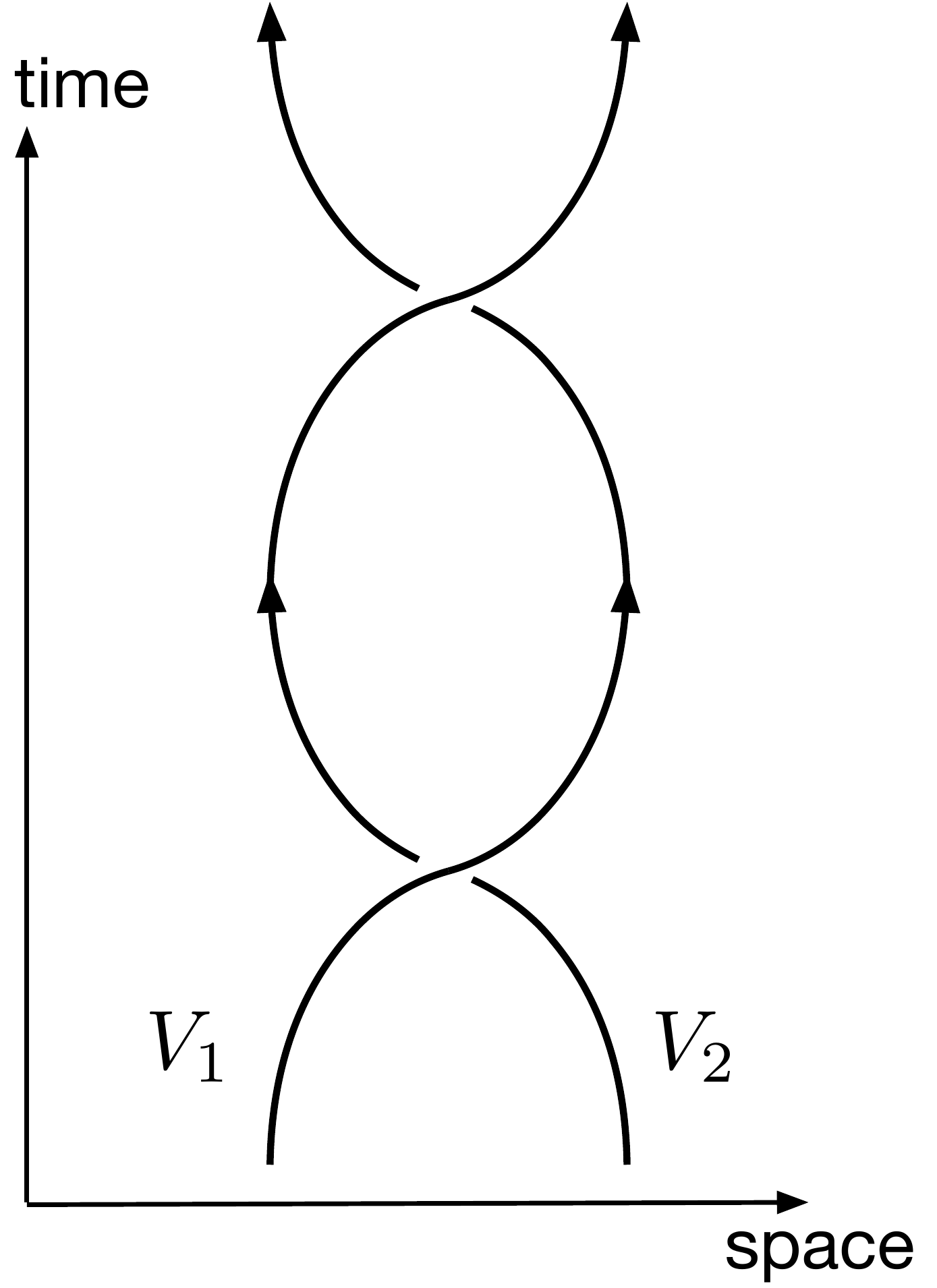}
\caption{ The world lines of two vortices rotating around each other in spacetime. The two vortices always keep finite spatial distance between them.  \label{fig:3}}
\end{figure}

Suppose that we have two vortices in spacetime, which we call $V_1$ and $V_2$, respectively.
$V_i$ is a magnetic brane for $\sphi_i$. By using the result \eqref{eq:QC3},
the electric charge $Q_{i, \sA }$ associated to $\sA$ on the vortex $V_i$ is given by
\beq
Q_{1, \sA} &= - \kappa \sphi_2 + \text{[const.]} \nonumber \\
Q_{2, \sA} &=  +\kappa \sphi_1 + \text{[const.]} 
\eeq
where the relative sign difference comes from the fact that 
\beq
\int \kappa\sphi_1  \d \sphi_2 \wedge \d \sA = -\int \kappa\sphi_2  \d \sphi_1 \wedge \d \sA.
\eeq
Now we rotate the two vortices around each other. The situation is analogous to Fig.~\ref{fig:1}, but in the present case we have two vortices instead of a monopole and an axion string.
Including the time direction, the spacetime world lines of two vortices are shown in Fig.~\ref{fig:3}.
The vortex $V_1$ goes around the vortex $V_2$ and hence the value of $\sphi_2$ on $V_1$
is changed as $\sphi_2 \to \sphi_2+1$ after one trip. The same is true for $V_2$ by reversing the indices $1 \leftrightarrow 2$.
Therefore, the electric charges on vortices are changed by the amount
\beq
\Delta Q_{1, \sA} &= - \kappa, \nonumber \\
\Delta Q_{2, \sA} &= + \kappa.
\eeq
Notice that we can always keep arbitrarily large distance between the two vortices, and we can do the process adiabatically 
so that no charged particles are created in the bulk $d$-dimensional spacetime. 
In fact, we can assume that charged particles in the $d$-dimensional bulk (but not on magnetic branes) are heavy enough
so that their production rate is extremely suppressed in adiabatic processes.
Thus there is no exchange of actual charged particles between 
the two vortices. Nevertheless, the charge on each of the vortices is changed. 

Locally on each vortex, it looks as if charge conservation is violated. However, the total charge $Q_{1, \sA} + Q_{2, \sA}$
is conserved and there is no inconsistency. The electric charge is teleported between the two vortices. 
Of course, this teleportation cannot violate causality. The time required for the vortices to go around each other and come back to the original position 
is greater than the time required for massless particles to travel between them. 
However, individual massless particles do not have electric charge 
if they are seen only as independent particles.
As a comparison, Coulomb force can transfer momenta (which are conserved quantities) between two systems keeping large distance.
One may interpret it as a transfer of momenta carried by virtual photons as suggested by Feynman diagrams. 
However, the charge teleportation is more subtle because
individual massless particles do not have electric charge. 
We emphasize again that there is no charged particle exchange between them.
If one wishes, we can put each vortex inside a shield which prohibits any particle penetration except for soft modes of massless gauge fields. 
The charge is still teleported even if there is such a shield.

\subsection{Charge teleportation in general dimensions}

Having discussed the simplest case in $d=3$, now we would like to discuss the general case with the action \eqref{eq:action}.
We consider two magnetic branes, one for $\sA^I$ and the other for $\sA^J$. We use capital letters $I, J$ to indicate that they are labels
associated to the two magnetic branes in spacetime.
We denote these magnetic branes as $M_I$ and $M_J$, 
respectively, and also by abuse of notation
we denote their world volumes by the same symbols $M_I$ and $M_J$, which have dimensions $d-p_I-2$ and $d-p_J-2$, respectively.
We consider a charge associated to the gauge field $\sA^k$ for a fixed $k$, and see that the charge is teleported between the two magnetic branes.

If $p_k > 1$, the meaning of ``charge'' associated to a higher form gauge field $\sA^k$ need to be carefully defined.
First, we consider the current of the theory on the magnetic brane $M_I$, denoted as $\sJ_{I,\, k}$, which is defined by the variation $\delta \sA^k$ as in Sec.~\ref{sec:QonBrane},
\beq
\int_{M_I} \delta \sA^k \wedge \sJ_{I,\, k} = \int \delta \sA^k \wedge \sJ_{I,\, k} \wedge \delta_{M_I}
\eeq
where $\delta_{M_I}$ is the delta function $(p_I +2) $-form localized on $M_I$.\footnote{Given a delta function $\delta_{M_I}$, 
we define the orientation of $M_I$ so that $\int_{M_I} \omega = \int \omega \wedge \delta_{M_I}$ holds
for any $\omega$.}
The current on $M_J$ is defined in the same way.

Next we choose a $(d-p_k)$-dimensional submanifold (including orientation) $ \Sigma$ on space (rather than spacetime),
and set $\Sigma_t = \{t \} \times \Sigma$ at a fixed time $t$.
Then we can consider the charge associated to this submanifold. 
We want to focus on the charges on the magnetic branes $M_I$ and $M_J$, so 
we take intersections of them with $\Sigma_t$,
\beq
\Sigma_{I,\, t} = M_I \cap \Sigma_t, \qquad \Sigma_{J,\, t} = M_J \cap \Sigma_t .
\eeq
The contributions to the charge from the theories on the magnetic branes $M_I$ and $M_J$ are
\beq
Q_{I,\, k}(\Sigma_{I,\, t}) = \int_{\Sigma_t} \sJ_{I,\, k} \wedge \delta_{M_I} =  \int_{\Sigma_{I,\, t}} \sJ_{I,\, k} \nonumber \\
Q_{J,\, k}(\Sigma_{J,\, t}) = \int_{\Sigma_t} \sJ_{J,\, k} \wedge \delta_{M_J} =  \int_{\Sigma_{J,\, t}} \sJ_{J,\, k}
\eeq
These equations specify the orientation of $\Sigma_{I,\, t} $ and $\Sigma_{J,\, t} $.

The point of the above definition is that we need to pick up a $\Sigma$ which is used for both $\Sigma_{I,\, t}$ and $\Sigma_{J,\, t}$.
If $\Sigma_{I,\, t}$ and $\Sigma_{J,\, t}$ were chosen independently on the respective magnetic branes, there would be no guarantee that the total charge $Q_{I,\, k}(\Sigma_{I,\, t})  +  Q_{J,\, k}(\Sigma_{J,\, t}) $
is conserved. Charge conservation is a statement that the integral of the total current $\cJ_k$ on a given $\Sigma$ is conserved.
This was not a problem for the case $p_k=1$ since in that case $\dim \Sigma = d-1$ and there is a unique nontrivial choice of $\Sigma$
in $(d-1)$-dimensional space (rather than spacetime).

By using the result \eqref{eq:QC3}, we get after slightly changing indices and using the relation $\kappa_{ikj }= (-1)^{(p_j+1)(p_k+1)}\kappa_{ijk }$ that
\beq
Q_{I,\, k}(\Sigma_{I,\, t}) & =  (-1)^{p_J (p_k+1) +1} \kappa_{IJk }  \int_{ \Sigma_{I,\, t}}    \sA^J  + \text{[const.]} \nonumber \\
 Q_{J,\, k}(\Sigma_{J,\, t}) & =   (-1)^{p_I (p_k+1)+1} \kappa_{JIk }  \int_{ \Sigma_{J,\, t}}    \sA^I  + \text{[const.]} , \label{eq:twoQ}
\eeq
where we have taken into account only the fields created by the magnetic branes labelled by $I$ and $J$, so there is no summation over indices $I$ and $J$. 
From these equations, we can already see that the presence of one magnetic brane affects the charge on the other,
because magnetic branes create nontrivial fields for the corresponding gauge fields.

Suppose that we move the two magnetic branes $M_I$ and $M_J$ while keeping finite distance between them.
We perform this process during a time interval $t \in [0, T]$,
and we assume that the system at the final time $t =T$ comes back to the original position as the one at the initial time $t =0$.
The world volumes $M_I$ and $M_J$ sweep submanifolds of spacetime between $t=0$ and $t=T$.
If $M_I$ and $M_J$ are vortices, the situation is as depicted in Fig.~\ref{fig:3}. 

\begin{figure}
\centering
\includegraphics[width=.8\textwidth]{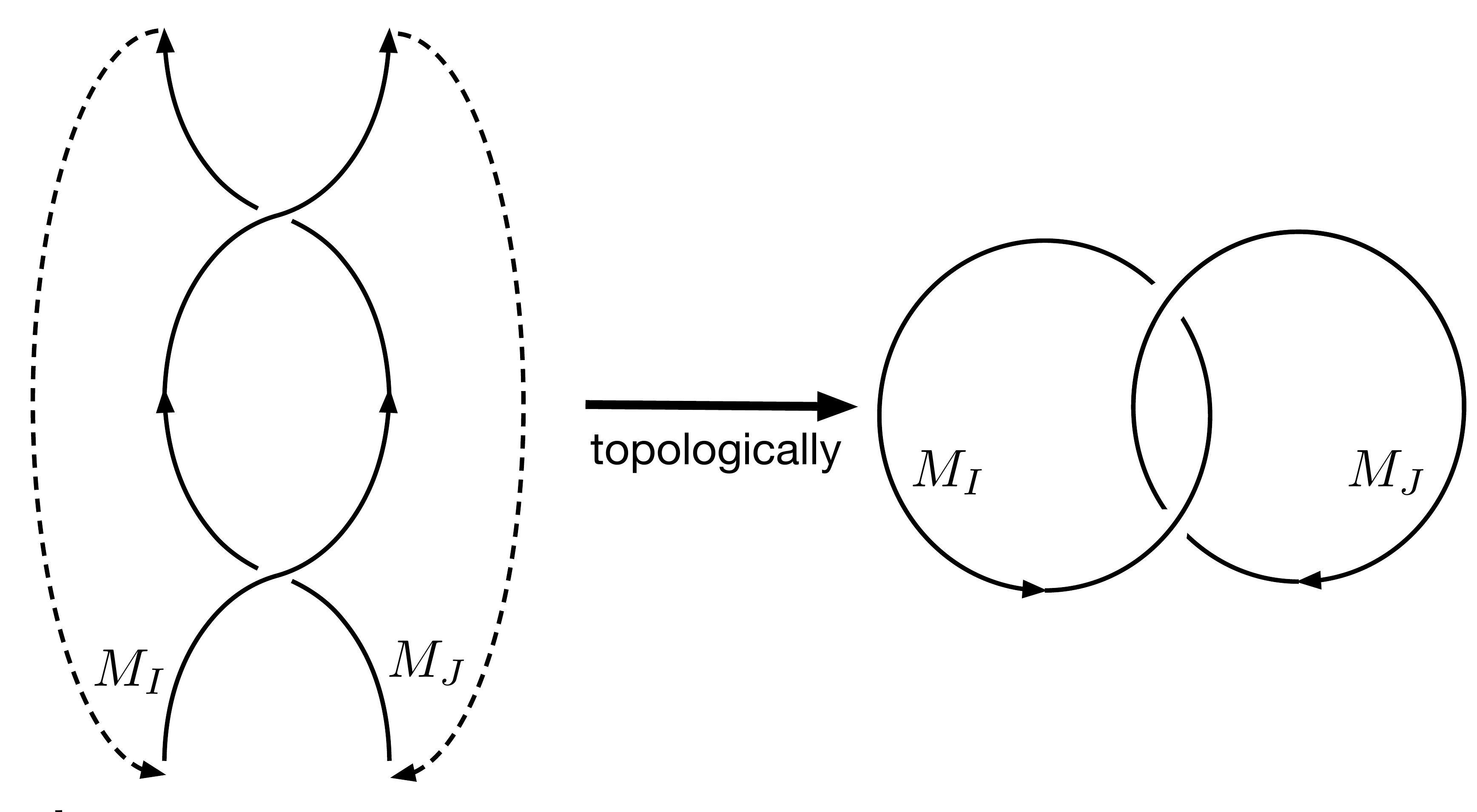}
\caption{  Left: By adding uphysical pieces to the world volumes represented by dashed lines, $M_I$ and $M_J$ form closed submanifolds.
Right: Two closed submanifolds $M_I$ and $M_J$ have a nontrivial linking. \label{fig:4}}
\end{figure}

Just for convenience of topological computations, we add additional pieces to the world volumes as follows.
After reaching $t =T$, we imagine that the branes ``go back in time'' from $t=T$ to $t=0$,
at fixed spatial positions. For example, in the case of a vortex in $d=3$,
we put a vortex at a fixed point in space and make it go from $t = T$ to $t =0$. 
See the left of Fig.~\ref{fig:4}.
Then the vortex forms a loop in spacetime,
by first going from $t =0$ to $t=T$ with varying spatial position and then going back in time from $t=T$ to $t=0$ at a fixed point.
More generally, the world volumes $M_I$ and $M_J$ are now closed submanifolds without boundary.
Alternatively, it is also possible to formally put a periodic boundary condition $t \sim t + T$ in the time coordinate without considering the pieces going back in time.
$M_I$ and $M_J$ are still closed.
In any case, the following conclusions are not changed.
We make $M_I$ and $M_J$ closed so that the amount of charge teleportation will be given by a topological invariant called the linking number.
See the right of Fig.~\ref{fig:4} for the situation.
But we remark that this is not essential at all.

The world volumes $M_I$ and $M_J$ are now closed submanifolds in spacetime.
Let us take intersections of them with $\bR \times \Sigma$, where $\bR$ is the time direction and $\Sigma$ is 
the submanifold in space introduced above to define a charge associated to $\sA^k$.
We denote them as $\Gamma_I$ and $\Gamma_J$ respectively,
\beq
\Gamma_I = (\bR \times \Sigma) \cap M_I, \nonumber \\
\Gamma_J =  (\bR \times \Sigma) \cap M_J
\eeq
Then the charges \eqref{eq:twoQ} are changed by the above process as
\beq
\Delta Q_{I,\, k} &=  (-1)^{p_J( p_k+1)+1} \kappa_{IJk }  \int_{ \Gamma_I}    \sF^J , \nonumber \\
\Delta Q_{J,\, k} &=  (-1)^{p_I( p_k+1)+1} \kappa_{JIk }  \int_{ \Gamma_J}    \sF^I . \label{eq:DQ}
\eeq
We will show that they are given by the linking number, and the total charge is conserved, $\Delta Q_{I,\, k} + \Delta Q_{J,\, k}=0$.

\begin{figure}
\centering
\includegraphics[width=1.0\textwidth]{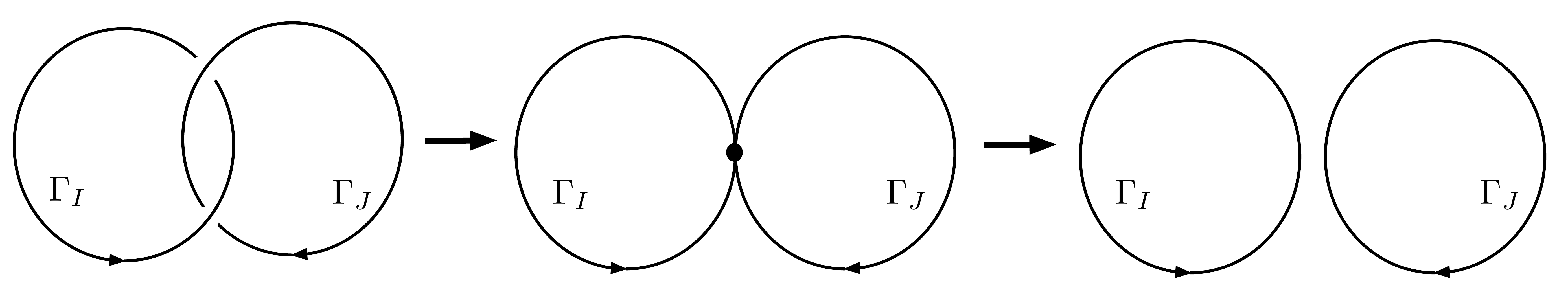}
\caption{ When we try to separate $\Gamma_I$ and $\Gamma_J$, they cross at a point as in the middle figure. \label{fig:5}}
\end{figure}

To understand the topological situation better, it is important to see the dimensions of $\Gamma_I$, $\Gamma_J$ and $\bR \times \Sigma$. They are given by
\beq
\dim \Gamma_I = p_J +1, \qquad \dim \Gamma_J = p_I +1, \qquad \dim( \bR \times \Sigma) = d- p_k +1.
\eeq
We also notice that the cubic Chern-Simons term \eqref{eq:action} requires $p_I +p_J + p_k +2 =d$. Therefore, the dimensions of these manifolds are related as
\beq
\dim( \bR \times \Sigma) = 1+ \dim \Gamma_I + \dim \Gamma_J . \label{eq:linkdim}
\eeq
Mathematically, these are the dimensions in which we can define the linking number between $\Gamma_I$ and $\Gamma_J$ in $\bR \times \Sigma$.
The linking number, up to sign, may be defined as follows. 
We try to separate two submanifolds $\Gamma_I$ and $\Gamma_J$ inside $\bR \times \Sigma$ by continuously deforming them away from each other.
Eventually the two submanifolds are far from each other. See Fig.~\ref{fig:5}.
For example, we may assume that $\Gamma_I$ is moved to $x = - \infty$ and $\Gamma_J$ is moved to $x = +\infty$ by the deformation,
where $x$ is a space coordinate. 
During the deformation process, $\Gamma_I$ and $\Gamma_J$ may cross at some points as in the middle of Fig.~\ref{fig:5}.
The fact that the crossings occur at points (rather than lines, surfaces, etc.) is due to the relation \eqref{eq:linkdim},
as least if the deformation is sufficiently generic.
Depending on the orientation of how the crossings occur, we assign a sign to each crossing point.
The sum of these signs over all crossing points is the definition of the linking number. 

For example, in the case of two vortices in $d=3$, we take $\bR \times \Sigma$ to be the entire spacetime.
The world line of each vortex is a loop as discussed above, and two loops inside a 3-dimensional manifold (such as $\bR \times \Sigma = \bR^3$) may be linked.
The nontrivial linking can be detected by counting the number of crossing points when we try to separate the two loops to a position in which they are not linked. 

We will show that $\Delta Q_{I,\, k}$ and $\Delta Q_{J,\, k}$ are given by the linking number between $\Gamma_I$ and $\Gamma_J$. 
This can be shown as follows.
Suppose that we deform $\Gamma_I$ to $\Gamma'_I$, and let $\h {\Gamma}_I$ be the manifold which is swept by the deformation process.
In other words, $\h {\Gamma}_I$ is a manifold which has boundaries $\partial \h {\Gamma}_I = \Gamma_I \sqcup \overline{\Gamma'_I}$, 
where the overline on $\overline{\Gamma'_I}$
means orientation reversal, and $\sqcup$ means disjoint union. 
We can assume that $\Gamma'_I$ which is obtained after the deformation is at infinity so that the integral of $\sF^J$ on $\Gamma'_I$ is zero,
\beq
 \int_{ \Gamma'_I}    \sF^j =0.
\eeq
Then by Stokes theorem, we get
\beq
\Delta Q_{I,\, k} &= (-1)^{p_J( p_k+1)+1}  \kappa_{IJk }  \int_{  \Gamma_I}    \sF^J -    (-1)^{p_J( p_k+1)+1}  \kappa_{IJk }  \int_{ \Gamma'_I}    \sF^J \nonumber \\
& =  (-1)^{p_J( p_k+1)+1}  \kappa_{IJk }  \int_{ \h \Gamma_I}   \d \sF^J = (-1)^{p_J( p_k+1) + 1 + p_J(d-p_J) } \kappa_{IJk } \int_{\h \Gamma_I} \delta_{M_J},
\eeq
where we have used \eqref{eq:magneticsign}.
When restricted to the submanifold $\bR \times \Sigma$, the delta function $\delta_{M_J}$ is equal to $\delta_{\Gamma_J}$, which
is the delta function Poincare-dual of $\Gamma_J$. More precisely, we define the orientation of $\Gamma_J$ as the Poincare dual of $\delta_{M_J}|_{\bR \times \Sigma} = \delta_{\Gamma_J}$.
We get 
\beq
\int_{\h \Gamma_I} \delta_{\Gamma_J}  = \int_{\bR \times \Sigma} \delta_{\Gamma_J} \wedge  \delta_{\h \Gamma_I}  =   \sum_{p \in \h \Gamma_I \cap \Gamma_J } \epsilon_p,
\eeq
where we have used the fact that $\h \Gamma_I \cap \Gamma_J $ consists of points in $\bR \times \Sigma$ if the deformation
$\h \Gamma_I $ is generic enough,
and $\epsilon_p = \pm 1$ is the sign determined by the orientation of the crossing between $\Gamma_J  $ and $\h \Gamma_I$
which we will explain in a moment. The sum on the right hand side is the definition of the linking number between $\Gamma_I$ and $\Gamma_J$.

The issue of orientation near a point $p$ can be studied locally. 
Suppose that we take a local coordinate system on $\bR \times \Sigma$, denoted as $y^1, \cdots, y^{d-p_k+1}$ such that $\Gamma_I$, $\Gamma_J$ and $\h \Gamma_I$ are given by
\beq
\Gamma_I &= \{ y^1 = c, \quad y^{p_J+3}= \cdots = y^{d-p_k+1}=0  \}, \nonumber \\
\h \Gamma_I &= \{ y^1 < c \quad y^{p_J+3}= \cdots = y^{d-p_k+1}=0  \}, \nonumber \\
\Gamma_J &= \{ y^1 = 0, \quad y^{2}= \cdots = y^{p_J+2}=0  \},
\eeq
where $c>0$ is a positive constant whose value is not important in the present topological discussions. The $\Gamma'_I$ is at $y^1 = - \infty$.
The point $p \in \h \Gamma_I \cap \Gamma_J$ is at $y^\mu = 0$ for all $\mu$.
We take coordinates $y$ including the orientation so that
\beq
\delta_{ \Gamma_I} &= \delta(y^{1}-c) \d y^1 \wedge  \delta(y^{p_J+3}) \d y^{p_J+3} \wedge \cdots \wedge \delta(y^{d-p_k+1}) \d y^{d-p_k+1} \nonumber \\
\delta_{\Gamma_J}  &= \delta(y^1) \d y^1 \wedge \delta(y^2) \d y^2 \wedge \cdots \wedge \delta(y^{p_J+2}) \d y^{p_J+2}.
\eeq
Notice that for a $(p_J+1)$-form $\omega$ we have
\beq
\int \omega \wedge \delta_{ \Gamma_I} = \int_{\Gamma_I} \omega = \int_{\h \Gamma_I} \d \omega = \int \d \omega \wedge  \delta_{ \h \Gamma_I} 
= (-1)^{p_J} \int  \omega \wedge \d \delta_{ \h \Gamma_I}.
\eeq
Thus we obtain
\beq
\delta_{\h \Gamma_I} &=(-1)^{p_J+1} \theta(c-y^1) \delta(y^{p_J+3}) \d y^{p_J+3} \wedge \cdots \wedge \delta(y^{d-p_k+1}) \d y^{d-p_k+1}
\eeq
and hence
\beq
 \epsilon_p = (-1)^{p_J+1} \int_{\bR \times \Sigma}   \delta(y^1) \d y^1 \wedge \cdots \wedge \delta(y^{d-p_k+1}) \d y^{d-p_k+1}.
\eeq
If the coordinate system $(y^1, \cdots, y^{d-p_k+1})$ has the same orientation as $\bR \times \Sigma$,
the integral gives $+1$. If it has the opposite orientation, the integral gives $ -1$.

Let us define
\beq
\sign(p) = \int_{\bR \times \Sigma}   \delta(y^1) \d y^1 \wedge \cdots \wedge \delta(y^{d-p_k+1}) \d y^{d-p_k+1}.
\eeq
In terms of it, we finally obtain
\beq
\Delta Q_{I,\, k} = (-1)^{ p_I p_J } \kappa_{IJk }  \sum_{p \in \h \Gamma_I \cap \Gamma_J } \sign(p).
\eeq
where we have used $d -p_J -p_k = p_I+2$ to simplify the sign factor.
This is the desired formula which gives $\Delta Q_{I,\, k}$ in terms of the linking number $ \sum_{p \in \h \Gamma_I \cap \Gamma_J } \sign(p)$.

Let us next consider $\Delta Q_{J,\, k}$. It can be computed by the same argument as above by exchanging $ I \leftrightarrow J$.
There is one-to-one correspondence between the points in the set $ \h \Gamma_I \cap \Gamma_J $
and the analogous set $ \h \Gamma_J \cap \Gamma_I $. To make it explicit, let us consider locally as above.
We introduce new coordinates 
\beq
&z^1 = c - y^1, \nonumber \\
& (z^2, \cdots, z^{p_I +2} ) = (y^{p_J+3}, \cdots, y^{d-p_k+1}), \nonumber \\
& (z^{p_I+3}, \cdots, z^{d-p_k+1}) = (y^2, \cdots, y^{p_ J +2} ).
\eeq
In terms of this new coordinate system, we have
\beq
\Gamma_J &= \{ z^1 = c, \quad z^{p_I+3}= \cdots = z^{d-p_k+1}=0  \}, \nonumber \\
\h \Gamma_J &= \{ z^1 < c \quad z^{p_I +3}= \cdots = z^{d-p_k+1}=0  \}, \nonumber \\
\Gamma_I &= \{ z^1 = 0, \quad z^{2}= \cdots = z^{p_I +2}=0  \}.
\eeq
The intersection $ \h \Gamma_J \cap \Gamma_I $ contains a point $q \in  \h \Gamma_J \cap \Gamma_I $
at $z^\mu=0$. The correspondence between $ \h \Gamma_I \cap \Gamma_J $ and $ \h \Gamma_J \cap \Gamma_I $ is made as $p \leftrightarrow q$. If $\h \Gamma_I$ and $\h \Gamma_J $
are generic enough, it is always possible to make such a correspondence. The reason is as follows. The intersection $\h \Gamma_I \cap \h \Gamma_J $
consists of line segments, and a single line segment has two boundary points $p \in \h \Gamma_I \cap \Gamma_J$ and $q \in  \h \Gamma_J \cap \Gamma_I$. Thus, by using these line segments,
we can make the one-to-one correspondence between points in $\h \Gamma_I \cap \Gamma_J$ and $\h \Gamma_J \cap \Gamma_I$. 
The relative orientation between the coordinate systems $y$ and $z$ is 
\beq
\d y^1 \wedge \cdots \wedge \d y^{d-p_k+1} = (-1)^{1 + (p_I+1)(p_J+1)} \d z^1 \wedge \cdots \wedge \d z^{d-p_k+1} 
\eeq
and hence we get 
$\sign(q) =  (-1)^{1 + (p_I+1)(p_J+1)} \sign(p)$.
Let us also recall that the Chern-Simons levels satisfy $\kappa_{IJk}= (-1)^{(p_I+1)(p_J+1)} \kappa_{IJk}$.
Therefore, we get
\beq
\kappa_{JIk} \sign(q) = - \kappa_{IJk} \sign(p).
\eeq

By using the above various signs, 
we can find the relation between $\Delta Q_{I,\, k}$ and $\Delta Q_{J,\, k}$,
\beq
\Delta Q_{J,\, k} &= (-1)^{p_I p_J } \kappa_{JIk }  \sum_{q \in \h \Gamma_J \cap \Gamma_I } \sign(q) 
= - (-1)^{p_I p_J  } \kappa_{IJk }  \sum_{p \in \h \Gamma_I \cap \Gamma_J } \sign(p) \nonumber\\
&= - \Delta Q_{I,\, k}.
\eeq
Therefore, we conclude that the total charge is conserved between the two systems. 
Each of $\Delta Q_{I,\, k}$ and $\Delta Q_{J,\, k}$ can be nonzero by considering a configuration with a nontrivial linking number.
Therefore, charges are teleported between the two magnetic branes. 
We have already discussed a concrete example in the Introduction and the previous subsection.

\subsection{Example: M5-branes in M-theory}
M-theory contains a 3-form field which we denote as $\sC$.
M2-branes have electric charges of $\sC$, while M5-branes have magnetic charges.
The 11-dimensional supergravity contains the Chern-Simons term
\beq
S \supset 2\pi \int \frac{1}{6} \sC \wedge \d \sC \wedge \d \sC.
\eeq
Therefore, we have all the necessary ingredient for the charge teleportation. 
If we move two M5-branes in some appropriate position, then M2-charges are teleported.
This is already established by the general result of the previous subsection.\footnote{
There is a somewhat analogous (although different) phenomenon in string theory, called the Hanany-Witten effect~\cite{Hanany:1996ie},
in which some branes are created by a reason of linking numbers.}

The anomalous degrees of freedom responsible for the charge teleportation is a chiral 2-form field on M5-branes.
It has a 2-form global symmetry~\cite{Gaiotto:2014kfa}, and can be coupled to a 3-form field which is $\sC$ in M-theory.
The chiral 2-form field has an anomaly of this higher form symmetry~\cite{Hsieh:2020jpj}, and this anomaly can create
M2-charges by the mechanism explained in Sec.~\ref{sec:charge}.

We can compactify M-theory to lower dimensions. 
For example, we can compactify it to 4 dimensions, and obtain monopoles and axion strings from M5-branes.
On the world sheets of axion strings, we get (chiral) scalars from the compactification of the chiral 2-form field.
On the world lines of monopoles, we have quantum mechanical models like \eqref{eq:QMmodel} obtained from the chiral 2-form.

Of course, charge teleportation is not limited to M-theory. 
For example, by compactifying M-theory on $S^1$, we get Type~IIA string theory.
By various string dualities, charge teleportation happens also in other string theories. 

\acknowledgments

HF is supported by the Director, Office of Science, Office of
High Energy Physics of the U.S. Department of Energy under the
Contract No. DE-AC02-05CH11231.
KY is partly supported by JSPS KAKENHI Grant-in-Aid (Wakate-B), No.17K14265.

\appendix

\section{$\U(1)$ charges in a 2-dimensional chiral fermion theory} \label{app:chiral}

In this appendix, we review the $\U(1)$ charge non-conservation in a 2-dimensional chiral fermion theory.
The argument based on anomalies is explained in Sec.~\ref{sec:charge}, and our purpose here is
to give a more explicit demonstration.
We consider a positive chirality fermion $\psi$ with $\U(1)$ charge 1. The $\psi$ has only a single component. 
The Lagrangian is
\beq
\cL = \i \psi^\dagger ( D_0 - D_1) \psi
\eeq
where $D_\mu = \partial_\mu - 2\pi \i \sA_\mu$ is the covariant derivative, and
we have normalized the gauge field $\sA_\mu$ according to our convention in the main text.
The time and space coordinates are denoted as $t = x^0$ and $\sigma = x^1$.

We consider this theory on a spatial manifold $S^1$ with radius $1$. (It is easy to do the following analysis for a general radius.)
We impose the anti-periodic boundary condition
\beq
\psi( t, \sigma + 2\pi  ) = - \psi(t, \sigma).
\eeq
This choice is not so essential for our purpose, but it slightly simplifies the later computations. 
We take the gauge field as
\beq
\sA_0=0, \quad \sA_1 = \frac{\sa}{2\pi },
\eeq
where $\sa$ is constant (or changes very slowly as a function of time $t$). In particular, we have
\beq
\oint_{S^1} \sA = \sa.
\eeq
Notice that $\sa$ defined by this integral over $S^1$ is gauge invariant up to large gauge transformations $\sa \to \sa + k$ for $k \in \bZ$,
which are realized by $\psi \to e^{\i  k \sigma} \psi$.

We can expand $\psi$ as
\beq
\psi(t,\sigma) = \frac{1}{\sqrt{2\pi } } \sum_{n \in \bZ + \frac{1}{2}} \psi_n \exp \left( -\i  (n + \sa) t - \i   n \sigma   \right).
\eeq
Then the canonical anticommutation relation is 
\beq
\{ \psi_n, \psi_m^\dagger \} = \delta_{n,m}, \quad \{\psi_n, \psi_m\}=0, \quad \{\psi_n^\dagger, \psi_m^\dagger \}=0.
\eeq
The Hamiltonian is
\beq
H = \sum_{n \in \bZ + \frac{1}{2}} (n + \sa) \psi_n^\dagger\psi_n + {\rm const.}
\eeq
The $\U(1)$ charge operator is given by
\beq
Q = \int \d \sigma \, \frac12 (\psi^\dagger \psi - \psi \psi^\dagger) = \sum_{n \in \bZ + \frac{1}{2}} ( \psi_n^\dagger \psi_n - \frac12 ),
\eeq
where we have defined it so that the quantization of a single pair $(\psi_n, \psi_n^\dagger)$ gives states of charge $\pm 1/2$.
This definition will make the charge finite in later computations. 

Now we can compute the $\U(1)$ charges of states in the Hilbert space. 
Suppose that a state $\ket{\Psi}$ satisfies 
\beq
( \psi_n^\dagger \psi_n - \frac12) \ket{\Psi} = \frac{1}{2} s_n \ket{\Psi}
\eeq
for each $n \in \bZ + \frac12$, where $s_n = \pm 1$. Then its charge is
\beq
Q =  \frac{1}{2} \sum_{n \in \bZ + \frac{1}{2}} s_n  e^{ - \epsilon |n + \sa | }
\eeq
where the factor $e^{ - \epsilon |n +\sa | }$ is put for regularization and we will take $\epsilon \to +0$.
Notice that $n + \sa$ are eigenvalues of the Dirac operator in the space direction $\i D_\sigma $
and hence this regularization of $Q$ is gauge invariant. 

For definiteness, let us assume that $- 1/2 < \sa < 1/2$ and compute the charge of the ground state.
For the ground state in the case $-1/2 < \sa < 1/2$, we have $s_n =-1$ for $n>0$ and $s_n = +1$ for $n<0$. Thus we get
\beq
Q &=  \frac{1}{2} \sum_{ k =0}^\infty  \left(  -e^{ - \epsilon (k +\frac12 +\sa ) } +   e^{ - \epsilon (k +\frac12 -\sa )} \right) 
= \frac{1}{2} \frac{\sinh(\epsilon \sa) }{ \sinh(\frac12 \epsilon)} \nonumber \\
& \xrightarrow{\epsilon \to 0}  \sa =  \int_{S^1} \sA.
\eeq
This depends continuously on $\sa$. Compare this result with \eqref{eq:QC3}.

If we change $\sa$ continuously from $\sa=0$ to $\sa =1$,
the state $\ket{\Psi}$ with $s_n = -1$ for $n>0$ and $s_n = +1$ for $n<0$ is no longer the ground state.
Notice that the gauge field configurations $\sa=0$ and $\sa=1$ are gauge equivalent as explained above. 
The state is adiabatically continued from the ground state at $\sa=0$
to an excited state at $\sa=1$. This is analogous to the case of the quantum mechanical model discussed in Sec.~\ref{sec:WQM}. 
In fact, the quantum mechanical model is obtained by bosonizing $\psi$ to a chiral boson and dimensionally reducing the chiral boson on $S^1$.


\bibliographystyle{JHEP}
\bibliography{ref}

\end{document}